\documentstyle[emulateapj]{article}

\begin{document}

\title{\bf Scattered Light Models of Protostellar Envelopes:\\
Multiple Outflow Cavities and Misaligned Circumstellar Disks}

\author{Kenneth Wood\altaffilmark{1, 2}, David Smith\altaffilmark{3},
Barbara Whitney\altaffilmark{4}, Keivan Stassun\altaffilmark{5}, 
Scott J. Kenyon\altaffilmark{1}, Michael J. Wolff\altaffilmark{4}, 
Karen~S.~Bjorkman\altaffilmark{6}}

\altaffiltext{1}{Harvard-Smithsonian Center for Astrophysics, 
60 Garden Street, Cambridge, MA 02138;
kwood@cfa.harvard.edu, kenyon@payne.harvard.edu}

\altaffiltext{1}{School of Physics \& Astronomy, University of St Andrews, 
North Haugh, St Andrews, The Kingdom of Fife, KY16 9AD, Scotland the Brave; 
kw25@st-andrews.ac.uk}

\altaffiltext{3}{Astronomy Department, 
The University of Texas, Austin, TX 78712;
dss@astro.as.utexas.edu}

\altaffiltext{4}{Space Science Institute, Suite 23, 1540 30th Street, 
Boulder, CO; bwhitney@colorado.edu, wolff@colorado.edu}

\altaffiltext{5}{Astronomy Department, University of Wisconsin, 475 N.
Charter Street, Madison, WI 53706; keivan@astro.wisc.edu}

\altaffiltext{6}{Ritter Observatory, Dept. of Physics \& Astronomy, 
University of Toledo, Toledo, OH 43606; karen@astro.utoledo.edu}

\begin{abstract}

Ground based imaging, imaging polarimetry, and recent 
{\it Hubble Space Telescope} WFPC2 and NICMOS images of 
protostars have revealed very complex scattered light patterns
that cannot be entirely explained by 2-D radiation transfer models.  
We present here for the first time 
radiation transfer models of T~Tau and IRAS~04016+2610 that are
fully 3-D, with the aim of
investigating the effects on image morphology of multiple illuminating 
sources and infalling envelopes that have been shaped by multiple 
outflows.  For T~Tau we have constructed scattered light 
models where the illumination of the surrounding envelope is by a binary with 
each source surrounded by its own small circumstellar disk or envelope.  
We find that the asymmetries 
in the WFPC2 image of T~Tau can be reproduced if the disks in the binary 
system are misaligned, consistent with a recently discovered bipolar outflow
believed to originate from the secondary.  
For IRAS~04016+2610 we find that the observed scattered light pattern 
can be reproduced by scattering in an envelope with cavities carved by 
two sets of bipolar outflows, suggestive of an embedded binary system.  

\end{abstract}


\section{Introduction}

In the standard picture of low mass star formation (e.g., Shu, Adams, 
\& Lizano 1987), Class~I protostars are young sources 
(age $\sim 10^5$ years) that are still surrounded by massive 
($\sim 0.01M_\odot$) circumstellar disks and partially obscured by their
natal envelopes. These stars appear highly extincted in the optical, 
with spectral energy distributions that peak in the mid- to far-infrared 
(e.g., Lada \& Wilking 1984; Beichman et al. 1986; Lada 1987; 
Myers et al. 1987; 
Wilking, Lada, \& Young 1989; Kenyon et al. 1990; Kenyon et al. 1993a), 
and exhibit highly polarized reflection nebulae 
(e.g., Whitney \& Hartmann 1993; Whitney, Kenyon, \& Gomez 1997; 
Lucas \& Roche 1997, 1998).  
From observations of Class~I protostars has emerged
a coherent picture of the protostellar environment. 
The basic model consists of a
dusty infalling envelope with cavities carved by the energetic 
outflows associated with this phase of star formation.  
The envelope absorbs the optical light and emits this radiation at 
longer wavelengths (e.g., Adams \& Shu 1986; Adams, Lada, \& Shu 1987; 
Kenyon et al. 1993a);  the reflection nebulae are formed by 
starlight scattering off the cavity walls (Bastien \& Menard 1990; 
Tamura et al. 1991; Whitney \& Hartmann 1993; 
Kenyon et al. 1993b; Whitney et al. 1997; Lucas \& Roche 1997, 1998; 
Lucas, Blundell, \& Roche 2000; Reipurth et al. 2000).  

Using this basic picture as a foundation, models have been constructed
including disks, envelopes, and bipolar cavities that can reproduce
in gross detail the observed spectral energy distributions and 
low-resolution ground-based imaging of many Class~I sources
(Whitney \& Hartmann 1992, 1993; Whitney et al. 1997; 
Fischer, Henning, \& Yorke 1996; 
Lucas \& Roche 1997, 1998; Stapelfeldt et al. 1998b).  
However, detailed agreement between these models and the observed
scattered light patterns of many Class~I sources has been less
forthcoming, in large part because the existing 2-D models have 
necessarily assumed axisymmetric circumstellar geometries.  
High-resolution HST images show that many sources exhibit very complicated 
scattered light patterns indicative of non-axisymmetric geometries and/or 
illumination (Burrows et al. 1996; 
Padgett et al. 1999; Stapelfeldt et al. 1995, 1998a,b, 1999; 
Terebey et al. 1998; Hartmann et al. 1999; Krist et al. 1998, 1999; 
Reipurth et al. 2000). 
Recent imaging polarimetry has provided additional compelling 
evidence that some sources exhibit departures from an axisymmetric 
disk-envelope-cavity system 
(Whitney et al. 1997; Lucas \& Roche 1997, 1998; Lucas et al. 2000).  

Two particularly good examples of sources with non-axisymmetric
circumstellar geometries are T~Tauri and IRAS~04016+2610. 
Ground based near-IR imaging polarimetry of 
IRAS~04016+2610 indicates {\it two} sets of bipolar outflow cavities 
(Lucas \& Roche 1997) and HST/NICMOS images show the complex scattered light 
pattern at higher resolution (Padgett et al. 1999).  
Ground based $J$, $H$, and $K$ images of T~Tau (Weintraub et al. 1992) 
were modeled by Whitney \& Hartmann (1993) with a 
standard envelope~+~bipolar cavity system, but the WFPC2 images 
(Stapelfeldt et al. 1998b) and near-IR adaptive optics images 
(Roddier et al. 1999, 2000) clearly show non-axisymmetric nebulosity.    

Motivated by these observed departures from axisymmetry, in this paper we 
develop scattered light models for IRAS~04016+2610 and T~Tauri 
that for the first time include 3-D 
geometries and illumination by multiple sources.  
We demonstrate that including the effects of multiple illumination 
sources, multiple outflow cavities, and mis-aligned circumstellar disks
can explain the observed scattered light patterns much more fully
than has been possible with 2-D modeling efforts hitherto. 
We have not undertaken an exhaustive search of parameter space when 
modeling the two objects as such investigations have been presented 
previously for axisymmetric models (Whitney \& Hartmann 1993; 
Kenyon et al. 1993b; 
Whitney et al. 1997).  For our models we adopt protostellar envelope 
parameters derived from previous 2-D modeling of scattered light images and 
spectral energy distributions of T~Tau and IRAS~04016+2610. The effects 
of changing the evnelope parameters (e.g., accretion rate, centrifugal 
radius, inclination) have been described in the aforementioned papers and 
our focus is on the departures from axisymmetry required to obtain a 
better match between the scattered light models and data.  Furthermore, 
since we view our models as a first step towards unravelling the complexities 
of the scattered light patterns, we have not attempted to quantify any 
figure of merit for model fits to the data.  Rather, the models we present 
are the first numerical simulations of published suggestions as to the 
origin of the asymmetric scattered light patterns of the two 
sources.  
In \S~2 we describe the scattered light images that form the basis 
for our modeling, 
in \S~3 we 
describe the ingredients of our models, in \S~4 we present the models for 
the individual sources, and we conclude in \S~5 with a discussion of our 
findings and further observational tests of our suggested geometries.

\section{Scattered Light Images}

The NICMOS data (F110W, F160W, F187W, F205W)
for IRAS~04016+2610 were presented by Padgett et al. 
(1999), and the WFPC2 data (F555W) on T~Tau by Stapelfeldt et al. (1998).  

The reduced and PSF-subtracted F555W image of T~Tau was provided to us by 
Karl Stapelfeldt.  
For IRAS~04016+2610 the NICMOS data were obtained from the HST archive and 
were re-reduced using the IRAF STSDAS package with the most recent set of 
reference files available. We created bad pixel masks using the
data-quality frames associated with each image, manually masking transient
bad pixels, and we also manually masked the bad column
128 (Bergeron \& Skinner 1997). The subsets of each 
image association were registered and combined using the IRAF STSDAS
pipeline {\sc calnicb} task. 

In order to subtract the point-spread function (PSF) 
from each of the re-calibrated HST images, we re-binned the images
by a factor of 2 and deconvolved them
using PSFs computed with TINYTIM version 4.4 (Krist 1995), which were
also oversampled by a factor of 2. The deconvolution was performed within
IRAF using the STSDAS {\sc lucy} task, which uses a Richardson-Lucy based
algorithm. The deconvolution process is an iterative one, and we stopped
the process after 5 iterations in the F110W filter, 10 iterations in
F160W, 15 iterations in F187W, 20 iterations in F205W. 

For IRAS~04016+2610 we also obtained an $I$ band image using the 
WIYN \footnote{The WIYN Observatory is a joint facility of the 
University of Wisconsin-Madison, Indiana University, Yale University, 
and the National Optical Astronomy Observatories.} 
telescope's CCD imager.  
The WIYN observations were obtained on 
13/14 December 1995 using the WIYN imager camera, as part of a 
feasibility study conducted under the WIYN queue observing 
program.  Conditions during the night were good, with some 
light cirrus and seeing of 0.8 - 1.4 arcsec.  Observations were 
made in each of the VRI bands, consisting of 3 - 300s 
exposures in each band.  Offsets of 10 arcsec were made between 
each exposure.  The data were reduced using the standard 
IRAF \footnote{IRAF is distributed by
the National Optical Astronomy Observatories, which are operated by
the Associations of Universities for Research in Astronomy, Inc.,
under cooperative agreement with the NSF.} 
data reduction packages for CCD images.  After reduction, 
the individual images for each exposure were combined into 
one image (900s total exposure) for each band.  Here we 
report only I-band results (the morphology seen at V and R 
is similar).  The combined I-band image is shown in Figure 4a.

\section{Model Ingredients}

In this section we present the basic ingredients of the 3-D radiation
transfer code used to model T~Tau and IRAS~04016+2610. The code is
comprehensive, and necessarily involves a large number of parameters
that in principle can be tuned to match observations. Wherever possible,
we constrain our model parameters based on known properties of the
sources. 

The basic geometry we consider consists of an infalling envelope with
bipolar cavities, illuminated from within by a single source or a 
binary system. For binary illumination, the two
components possess circumstellar disks that are not necessarily 
aligned with one another or with the envelope's rotation axis. 

\subsection{Envelope Geometry}

Of the two sources we model in this paper, T~Tau is a known binary (Dyck, 
Simon, \& Zuckerman 1982) 
and IRAS~04016+2610 shows evidence for binarity 
with possibly two bipolar outflow cavities (Lucas \& Roche 1997).  
Despite the binary nature of the sources, 
we adopt the rotational collapse geometry of Terebey, Shu, \& Cassen (1984,
hereafter TSC) developed for single star systems.  We thus in effect
assume that the circumbinary density is given by the TSC solution and 
neglect the effects of binary motion on the envelope structure.  The focus 
of our present study is the 3-D nature of the cavities and envelope 
illumination; future studies may include circumbinary density distributions 
predicted from binary star formation models.  

The TSC envelope density used in our simulations is given by
\begin{equation} \rho = \frac{\dot M}{4\pi}\left(\frac{GM} {
r_c^3}\right)^{-1/2} \left(\frac{r}{r_c}\right)^{-3/2}
\left(1+\frac{\mu}{\mu_0}\right)^{-1/2}\left(\frac{\mu}{\mu_0} +
\frac{2\mu_0^2 r_c} {r}\right)^{-1} \; ,
\end{equation}
where $\dot M$ is the mass
infall rate, $r_c$ is the centrifugal radius, $\mu = \cos\theta$ and 
$\mu_0=\mu(r\rightarrow \infty)$ is determined by 
\begin{equation} \mu_0^3 +
\mu_0 (r/r_c - 1) - \mu (r/r_c) = 0 \; .
\end{equation}
For our models, we vary $\dot M$ and $r_c$, in the range 
$\dot M = (2-10) \times 10^{-6}M_\odot$yr$^{-1}$ 
and $r_c = 50-100$AU (Strom 1994; Kenyon et al. 1993b; 
Whitney, et al. 1997).  Our radiation transfer code uses a 
linear Cartesian grid (\S~3.5) and in order to resolve the TSC density 
structure we adopt an outer radius for the envelope of 1500AU.  
We note that in the models of 
Whitney \& Hartmann (1993) and Whitney et al. (1997), outer radii of around 
10000AU were adopted.  We have found that while our smaller envelopes can 
reproduce the overall morphology of the scattered light images, larger 
envelopes produce redder colors (see the IRAS~04016+2610 models in \S4.2).

\subsection{Cavity Geometries} 
We explore three dimensional geometries by carving cavities in the 
axisymmetric 
TSC envelope.  By virtue of being 2-D, previous modeling 
(e.g., Whitney \& Hartmann 1993; Whitney et al. 1997; Lucas \& Roche 1997; 
Stapelfeldt et al. 1998) used 
bipolar cavities that were aligned with the envelope's rotational axis.  
In our 
models we adopt the same cavity shapes as used previously, but now allow the 
cavities to be offset from the envelope's 
symmetry axis and also allow for the existence of multiple sets of cavities,
as might result in a binary system. 

We consider three cavity shapes in our modeling: streamline, cylindrical, 
and curved.  In the streamline cavity shape, the cavity walls trace out
the trajectory of an infalling particle as given in the TSC envelope
solution. This cavity shape is approximately conical
at large radii, but at small radii becomes more curved.  The curvature 
increases the cross-section of the cavity walls exposed to direct starlight 
and results in more scattering than a purely conical shape and consequently 
a brighter cavity (see Whitney \& Hartmann 1993).  Cylindrical
cavities have straight walls parallel to the outflow axis.  
Curved cavities take the form $z = a \omega^b$ 
(e.g., Stapelfeldt et al. 1998b), 
where $\omega=\sqrt{x^2 + y^2}$ and $a$ and $b$ are 
parameters that can be adjusted to provide a good match to the data or 
to approximate an opening angle of a molecular outflow.  
We also allow for the presence of dust within the cavities.

\subsection{Illumination}

As stated earlier, T~Tau is a binary and there are indications that 
IRAS~04016+2610 may also be a binary.  
Therefore we assume 
the illumination is by binary star~+~disk systems as used by 
Wood, Crosas, \& Ghez (1999) in their scattered light model of GG~Tau's 
circumbinary disk.  In this model we assumed two point sources of 
illumination, but with the radiation angular distribution of a 
star~+~disk system.  In this approximation we do 
not explicitly treat the radiation transfer through 
the circumstellar disks, but assign an angular distribution for the 
flux emerging from the star+disk systems.  The angular dependence of the 
illumination of these ``star~+~disk'' systems is displayed in Figure~1, 
showing the small emergent flux level at high inclinations due to the 
dense equatorial disk (Whitney \& Hartmann 1992).  
The angular distribution of the illuminating sources 
was set up so that the sources would be at the center of the cavities, with 
the polar direction of the illumination pattern aligned with the cavity axis.

\subsection{Dust Opacity and Scattering Parameters}

As with previous modeling of Taurus protostars (Kenyon et al. 1993b; 
Whitney et al. 1997; Wood et al. 1998) we adopt the 
albedo, opacity, and scattering parameters 
for a Kim, Martin, \& Hendry (1994, hereafter KMH) grain mixture, typical 
of grains in the interstellar medium (ISM).  The scattering phase 
function is assumed to be a Henyey-Greenstein phase function 
(Henyey \& Greenstein 1941) with asymmetry 
parameter $g$.  
In more evolved disk systems the spectral index and level of the long 
wavelength continuum emission provides strong evidence for dust grains 
that are larger than ISM grains (Beckwith et al. 1990; 
Beckwith \& Sargent 1991).  However, Whitney et al. (1997) found that the 
KMH parameters gave a good match to the 
colors and sizes of images Class~I sources, but that they underestimated the 
polarization level.  The KMH wavelength dependent opacity and scattering 
parameters we adopt in this paper are presented in Table~1.

\subsection{Radiation Transfer}

We perform the radiation transfer with a Monte Carlo 
code that allows for arbitrary sources of emission and multiple scattering 
within an arbitrary geometry.  The scattering code is based on that described 
by Code \& Whitney (1995), but has now been modified to run on a three 
dimensional linear Cartesian grid (Wood \& Reynolds 1999) and includes  
forced first scattering (Witt 1977) and a ``peeling off'' procedure 
(Yusef-Zadeh, Morris, \& White 1984).  These modifications enable us to 
construct model images of three dimensional systems from specified viewing 
angles very efficiently.  We have tested our 3D code against results from 
the axisymmetric code of Whitney \& Hartmann (1993), finding 
excellent agreement between the two codes.  

The output of each Monte Carlo simulation is a two dimensional image 
including the spatially resolved polarization (Stokes $I$, $Q$, and $U$ 
images).  In order to compare our models with the observations we convolve 
our high resolution models with appropriate PSFs using the TINYTIM 
software (Krist 1995).  
From these images we calculate the integrated polarization 
and by combining simulations at different wavelengths we calculate 
colors for comparison with observations.  
The colors we compute are subject to 
a normalization set by the intrinsic color of the illuminating source(s).  
The observed color is $J-H = (J'-H')+(J_0-H_0)$, where 
$J_0-H_0$ is the source color and $J'-H'$ is the color of the Monte Carlo 
simulation.  In their simulations, Whitney et al. (1997) adopted source 
colors for a typical M0 T~Tauri star plus reprocessing disk from 
Kenyon et al. (1994), $J_0-H_0 = 0.77$ and $H_0-K_0=0.43$.  In our 
models of IRAS~04016+2610 we adopt the same source colors.  However, for 
T~Tau the spectral type is K0 or earlier and since we are modeling only 
the WFPC2 F555W image the intrinsic source colors are not required.

\section{Results}

In this section we present our models for T~Tau and IRAS~04016+2610.  
As our aim in this paper is
to demonstrate the usefulness of our 3-D 
modeling method, we report here on those models that best reproduce
the observed scattered light images, and discuss the salient properties
of those models. As we discuss in detail for each source, we 
constrain the parameters of our models based on the 
available data from the literature. 
Nonetheless, the uniqueness of the models presented is not demonstrated.
These models should be regarded as
existence proofs and as suggestions for more detailed 3-D investigations
of these and other Class~I sources in the future.

\subsection{T Tau}

\subsubsection{Previous 2-D models and evidence for source asymmetry}

T Tau is a binary star with one component seen directly and the secondary 
($0.7$ arcsec distant) only detected at IR and longer wavelengths (Dyck, 
Simon, \& Zuckerman 1982; Ghez et al. 1991).  
The bolometric luminosities of the sources are estimated to be $9L_\odot$ 
for the primary (Ghez et al. 1991) and up to $14 L_\odot$ for the 
secondary (Cohen,
Emerson, \& Beichman 1989) .  Therefore, although undetected at optical 
wavelengths, the secondary may 
be emitting a lot of radiation which we only observe when it 
is scattered into our sightline.  
The optical light of the secondary may be obscured by either its 
location within 
a dense circumbinary envelope, or occulted by its own circumstellar disk 
edge-on to our line of sight.  
The WFPC2 scattered light pattern shows departures from axisymmetry 
with one side of the cavity appearing brighter than the other 
(Figure 2a; Stapelfeldt et al. 
1998b).  A similar asymmetric scattered light pattern is seen in the near-IR 
adaptive optics images of Roddier et al. (1999).

The radial distribution of the near-IR scattered light surrounding T~Tau 
(Weintraub et al. 1992) was modeled by 
Whitney \& Hartmann (1993) adopting a near pole-on orientation of a 
TSC envelope ($\dot M = 2\times 10^{-6}M_\odot$yr$^{-1}$, $r_c = 100$AU) 
and evacuated bipolar cavities.  They found that cavities with streamline 
shapes give too steep a gradient in the radial distribution of the 
scattered light and suggested that curved cavities could provide a 
better fit to the data.  Stapelfeldt et al. (1998b) modeled T~Tau's 
scattered light pattern assuming single scattering in a TSC envelope and 
investigated what viewing angle and cavity curvature is required to match 
their HST images.  They estimated the inclination of the system 
to be in the range $i=25^\circ$ to $45^\circ$ with curved cavities 
($z = a r^b$, $a = 5\times 10^{-3}$, $b=1.95$).
Figure 2b shows the axisymmetric single source model of Stapelfeldt et al. 
(1998b). The scattered light distribution in the model is 
necessarily symmetric,
and does not reproduce the strong asymmetry observed in the WFPC2 image.

\subsubsection{New 3-D model}

A binary system with shadowing caused by misaligned circumstellar 
disks may yield a scattered light pattern such as that observed
in T~Tau.  Indeed, Stapelfeldt et al. 
(1998b) suggested that the unseen secondary may play an important role in 
forming the scattered light pattern.  We have therefore 
constructed a scattered light model for T~Tau incorporating illumination 
from a binary star system and allowed the alignment of the circumstellar 
disks to depart from coplanarity.

Our modeling of T~Tau begins with the axisymmetric TSC envelope model 
of Whitney \& Hartmann (1993) with the curved cavities derived by Stapelfeldt 
et al. (1998b), and we adopt $i=40^\circ$ which results in a visual 
extinction through the envelope to the primary of $A_V\approx 2$, close to 
the observed value. However,
in our model the axisymmetry is broken by the illuminating sources.  
The illumination is by a binary star~+~disk system where the primary source 
is aligned with the rotational axis of the envelope and the secondary 
(larger bolometric 
luminosity) has its disk aligned so that its polar axis points towards 
one side of the cavity wall.  The direction of the secondary's axis is
chosen to agree with 
the large outflow discovered by Reipurth, Bally, \& Devine (1997), which is 
believed to originate from the secondary. 

We assume the secondary emits 1.5 times more 
photons than the primary, based on  their estimated bolometric luminosity 
ratio.  The secondary is located 90AU from the primary in the envelope's 
equatorial plane 
with its disk rotational axis at $\phi = -120^\circ$, 
$\theta = 70^\circ$ (where 
$\theta$ and $\phi$ are the usual polar coordinates).  
For our adopted viewing angle the secondary disk is edge-on to our 
line of sight and consequently its direct light is highly extincted at 
optical wavelengths, consistent with its very red spectral energy 
diistribution.  Furthermore the disk orientation 
blocks the light from the secondary from reaching one side of the cavity.  
In this model we assume the primary and secondary are within the same cavity, 
created by a combination of outflows from both sources.

In Figures 2c and 2d we show new 3-D scattered light models that
better reproduce the observed scattered light image of T~Tau.
We rotated these model 
images by $20^\circ$ to allow for comparison with the WFPC2 data.  
The model in Fig.~2c shows the scattered 
light pattern when the illumination is by misaligned circumstellar disks 
where the angular dependence of the illumination for each star~+~disk system 
is that of the WH92 model in Fig.~1.  
The non-axisymmetric illumination yields an 
asymmetric nebula with one side brighter than the other, similar to what 
is observed.  Direct light from 
the secondary is blocked due to our adopted edge-on orientation of its 
circumstellar disk, consistent with the lack of observed optical light
from the secondary. 

In Fig.~2d we allowed the secondary's 
illumination pattern to depart from that of the WH92 star~+~disk of Fig.~1 in 
order to investigate whether we could increase the asymmetry in our 
scattered light models.  In this model, the illumination pattern from 
the secondary is 
more concentrated towards the polar regions (see Fig.~1).  
Such an illumination 
may arise if the secondary's circumstellar material is arranged in a highly 
flared disk (e.g., Chiang \& Goldreich 1999) or a small envelope 
(Roddier et al. 2000).  With the 
more extreme illumination pattern, the asymmetry of the model scattered 
light nebula is enhanced.

\subsubsection{Efficacy and limitations of 3-D model}

Our model for T~Tau resembles the WFPC2 image in that 
it reproduces the asymmetry in the scattered light nebula.  Support for 
our model may be found in observations which show a very complex pattern 
of outflows in CO (Schuster, Harris, \& Russell 1997; Leverault 1998) and 
molecular hydrogen (Herbst, et al. 1996; van Langevelde et al. 1994).  As 
with IRAS~04016+2610 (see below), the plethora of emission knots and 
filaments is suggestive of a precessing jet and/or multiple outflow sources.  
The most compelling support for the orientation of the secondary in our 
models is that its disk axis is aligned with the direction of the outflow 
discovered by Reipurth et al. (1997) and attributed to the secondary.

In our model the circumstellar 
disk axes are offset from one another, the primary viewed at a low 
inclination and the secondary oriented close to edge-on to our line 
of sight.  This is not an unreasonable assumption, given the fact that 
the secondary in the T~Tau system is undetected at optical wavelengths.  
Also, a system with 
such a geometry has been observed in the binary HK~Tau where the 
primary is seen directly and the secondary's scattered light nebula 
indicates a highly inclined circumstellar disk (Stapelfeldt et al. 1998a; 
Koresko 1998).

The inclusion of a secondary source not only breaks the asymmetry in the 
scattered light image, but also increases the contrast between the bright 
cavity walls and the region interior to this.  In their axisymmetric 
modeling, Stapelfeldt et al. (1998b) produced a contrast 
no greater than about two 
between the nebular arc and the inner region, whereas the WFPC2 image shows 
a contrast in excess of four.  Our models can produce contrasts in excess 
of four as shown in Fig.~3.  This figure shows horizontal intensity 
cuts at various offsets across the axisymmetric model (Fig.~2b) and our 
new model (Fig.~2d).  The larger brightness contrast of our 
non-axisymmetric model (solid line) compared to the axisymmetric model 
(dashed line) is clearly evident.  

While the overall agreement between 
our $V$ model and the WFPC2 image is very good, we cannot reproduce the 
thickness of T~Tau's nebular arc.  This discrepancy is also present in 
the model of Stapelfeldt et al. (1998b).  
In our model we have assumed that T~Tau's IR 
companion is a disk source seen almost edge-on.  However, as it is not 
detected in the optical, but bright in the IR, we may be viewing at 
grazing incidence through the upper layers of a flared disk or envelope.  
As such, detailed 3D modeling of the spectral energy distribution will 
help in refining our model.  We are currently developing techniques for 
such 3D radiative equilibrium modeling within our Monte Carlo codes 
(Bjorkman \& Wood 2001).

\subsection{IRAS 04016+2610}

\subsubsection{Previous 2-D models and evidence for source asymmetry}

IRAS~04016+2610 is a Taurus Class I source with a bolometric
luminosity of $3.7 L_\odot$ (Kenyon \& Hartmann 1995) and displays 
a very complex scattered light pattern (Whitney et al. 1997; 
Lucas \& Roche 1997; Padgett et al. 1999; Lucas et al. 2000).  
The HH knots mapped by Gomez et al. (1997) appear scattered throughout the 
environs, indicative of a precessing jet or multiple outflows.  Evidence 
in favor of multiple 
outflows is found in the C$^{18}$O molecular line profiles of Zhou et al. 
(1994) that show multiple peaks both red and blueshifted.  Perhaps the most 
compelling evidence of multiple outflows is the near-IR imaging polarimetry 
presented by Lucas \& Roche (1997, Fig.~11) and 
Whitney et al. (1997, Fig.~3).  
Lucas \& Roche presented contours of polarized flux (essentially mapping the 
scattered light) which they interpreted as arising from scattering in an 
envelope with {\it two} sets of dust-filled bipolar outflow cavities with 
projected axes that are almost perpendicular to one another.  Our 
WIYN $I$ band image (shown in Fig.~4a) displays bright elongated limbs that 
are aligned with the directions of the suspected outflow cavities mapped by 
Lucas \& Roche (1997).  

In previous axisymmetric modeling of IRAS~04016+2610, Kenyon et al. 
(1993a,b) obtained a good match to the spectral energy distribution 
and low resolution near-IR imagery adopting a TSC envelope 
with bipolar streamline cavities viewed at $i= 60^\circ$.  Whitney 
et al. (1997) adopted a similar model 
($\dot M = 5\times 10^{-6}M_\odot$yr$^{-1}$, $r_c = 50$AU) for their near-IR 
imaging polarimetry and found that while it yielded colors close to those 
observed, to match the observed nebular pattern would 
require a fully 3D model (as further demonstrated by Lucas \& Roche 1997; 
Lucas et al. 2000).  A single wide outflow cavity cannot reproduce the 
distinct V-shape in the WIYN image and imaging polarimetry.  A single cavity 
results in a conical scattered light image, but cannot reproduce the large 
brightness contrast between the arms of the V-shape and the interior region, 
e.g., see the axisymmetric T Tau model (Fig.~2b) and also Whitney \& Hartmann 
(1993) and Stapelfeldt et al. (1998b).

\subsubsection{New 3-D model}

In our 3D modeling of IRAS~04016+2610 we adopt the TSC envelope geometry 
and test the multiple cavity scenario proposed by Lucas \& Roche (1997).  
The TSC envelope has $\dot M = 5\times 10^{-6}M_\odot$yr$^{-1}$, 
$r_c = 50$AU (Whitney et al. 1997).  
Our scattered light model shown in Fig.~4 
has two sets of bipolar cylindrical, dust-filled cavities carved in the 
TSC envelope.  The primary cavity (aligned along the rotation axis of the 
TSC envelope) has a radius of 50AU and the secondary cavity has a radius 
of 80AU.  The viewing angle is $i=50^\circ$ relative to the primary cavity.  
Both cavities have a $J$ band optical depth of 2 measured from 
the center of the envelope along the cavity axis.  
To match the bright, elongated limbs in the WIYN image, 
we rotated the axis of the second bipolar cavity to lie along 
$\phi = -85^\circ$, $\theta = 80^\circ$ (see Fig.~5).  In this model the 
two bipolar cavities are almost perpendicular as noted by Lucas \& Roche 
(1997).  The scattered light images have been 
rotated clockwise by $220^\circ$ so that the scattered light pattern formed 
by the primary cavity (aligned with the TSC rotation axis) is aligned with 
the primary outflow cavity identified by Lucas \& Roche (1997), see the 
labeling in Lucas et al. (2000, Fig.~2).  

Lucas \& Roche (1997) interpreted their data as possible evidence for 
a binary carving two outflow cavities.  As such, we have adopted an 
unresolved binary for the illuminating sources.  The NICMOS images 
of Padgett et al. (1999) and centimeter observations of Lucas et al. (2000) 
did not resolve a binary companion.  We also constucted a model with the 
two outflow cavities described above, but illumination from a single source.  
Because the disk sources depart from isotropy only along equatorial 
directions (Fig.~1), we found that the scattered light pattern from a single 
source is not distinguished from that of the unresolved binary model 
of Fig.~4.  Therefore, unlike our T Tau model, we infer that for 
IRAS~04016+2610 it is the 
envelope structure that is most important in producing the image morphology 
and not asymmetric illumination.

\subsubsection{Efficacy and limitations of 3-D model}

Our model reproduces the overall size and morphology of the WIYN 
and NICMOS images very well.  
In the $I$ band model, the envelope is optically thick and it is the cavities 
that dominate the image morphology.  However, at optically thinner longer 
wavelengths the envelope structure is important in shaping the resulting 
image.  Our model has a single TSC envelope with the primary cavity along 
the rotation axis and the second cavity almost perpendicular to this.  If 
the source is indeed a binary then the envelope geometry is likely different 
from TSC.  Indeed, recent modeling of imaging spectroscopy by Hogerheijde 
(2001) adopted a 2000AU radius axisymmetric rotating disklike structure.  
Such a ``disk'' should not be confused with the smaller circumstellar disks 
we adopt and are seen in other Classical T Tauri stars (e.g., 
Burrows et al. 1996; Padgett et al. 1999).  The TSC rotational collapse 
geometry produces a large equatorial density enhancement which may 
correspond to Hogerheijde's parameterized equatorial structure.  As such, 
the equatorial enhancement of TSC and Hogerheijde's ``disk'' represent 
a different evolutionary stage from disks around Classical T~Tauri stars 
(Shu et al. 1987).  A more detailed model of IRAS~04016+2610 than we have 
undertaken should simultaneously address the envelope geometry and 
velocity structure.  
Nevertheless, our models which incorporate the TSC parameters determined 
by Kenyon et al. (1993b) and Whitney et al. (1997) 
provide support for Lucas \& Roche's (1997) dual bipolar cavity 
interpretation of their imaging polarimetry.

Our model colors are $J'-H'=1.34$, $H'-K'=0.79$ giving actual colors of 
$J-H=2.11$, $H-K=1.22$ when combined with the adopted intrinsic colors 
(\S~2.5).  The observed colors 
for IRAS~04016+2610 are $J-H=1.95$, $H-K=2.02$ (Kenyon \& Hartman 1995; 
Whitney et al. 1997).  The failure of both our models and those of Whitney 
et al. (1997) to reproduce these colors (both models predict 
$[J-H]>[H-K]$) may indicate that the intrinsic 
source colors are different than those we have adopted.  

While we have been succesful at reproducing the image morphology, our 
models suffer from the same problem as those of Whitney et al. (1997) 
in that they predict lower polarizations than observed.  Whitney et al. 
attribute this to the low intrinsic polarization of the KMH grains.  
In our case, the resolved polarization is lower than observed due to the 
dust grains, but we also find much lower levels for the total unresolved
polarization in our models.  We typically predict an integrated polarization 
of around 3\% in the near-IR, whereas the observed polarization ranges from 
$P_J=21\%$ to $P_K=8\%$ (Whitney et al. 1997).  Another problem 
is that neither our models or those of Whitney et al. (1997) can reproduce 
the observed pattern of polarization vectors close to the source(s).  
So, while the dual cavity scenario proposed by Lucas \& Roche (1997) gives 
a good match to the overall image morphology, its failure to reproduce the 
observed polarization level and pattern in IRAS~04016+2610 indicates that 
the details of the model need to be refined.  In particular, polarization 
due to scattering off aligned grains can yield larger polarization 
values and we are investigating grain alignment in protostars to 
reconcile models with the large observed polarizations (Whitney \& Wolff 
2001, in preparation).

\section{Summary of Results}

We have constructed 3D models for the scattered light nebulae associated 
with IRAS~04016+2610, and T~Tau.  Our models assumed an axisymmetric infalling 
circumstellar envelope, but the axisymmetry was broken by multiple 
outflow cavities and non-axisymmetric illumination from binary stars.  

The asymmetry seen in the nebula surrounding T~Tau was modeled by 
introducing a secondary star+disk system with its rotation axis misaligned 
from the axis of the TSC envelope.  The orientation of the secondary was 
chosen so that its disk was oriented almost edge-on to our line of sight,
and was selected to be 
%The orientation of our model disk is 
perpendicular to the Herbig-Haro 
outflow attributed to the secondary (Reipurth et al. 1997).
This orientation results in the secondary being undetected in our scattered 
light models as its disk extincts the direct optical light.  

We reproduced the scattered light pattern associated with IRAS~04016+2610 
with a model which had two sets of bipolar outflow cavities in the 
TSC envelope.  This model affirms the notion of Lucas \& Roche (1997), 
who interpreted their near-IR imaging polarimetry as arising from scattering 
in an envelope with orthogonal bipolar cavities.  Further evidence for 
multiple cavities in the envelope is found in the complex pattern of 
outflow emission knots surrounding this source (Gomez et al. 1997).  
Our model fails, however, to reproduce the level and pattern of the 
near-IR polarization and we suspect that a more accurate treatment of 
polarization from scattering off aligned grains may provide a better 
match to the observations (Whitney \& Wolff 2001, in preparation).

%We have adopted a dust grain model in which the average grain size 
%is larger than that in the diffuse ISM and the dust opacity has a shallower 
%wavelength dependence than the ISM dust opacity.  This is in line with many 
%other studies of dust size and opacity in protostellar environments.

In the models presented in this paper, we adopted the axisymmetric 
TSC solution for the circumstellar envelopes.  This may not 
be appropriate as T~Tau is a known binary and IRAS~04016+2610 
shows strong evidence for binarity.  Also, other studies indicate that 
protostellar environments have material arranged with a shallower radial 
gradient than TSC (e.g., Barsony \& Chandler 1993).  
The extension of our Monte Carlo 
scattering codes to three dimensions now enables us to incorporate more 
complex circumstellar envelope structures, such as those predicted from 
binary star formation models.  However, this is among the first 
modeling investigations to explore the effects of three dimensional cavity 
structures and illuminations and has proved successful in reproducing the 
scattered light morphologies and in testing several scenarios proposed by 
others to explain the observations.  
While we have been fairly succesful in reproducing the scattered light 
morphologies, spectral energy distribution modeling incorporating the 
3D structures will provide a crucial test of our models.  This requires 
current axisymmetric radiation transfer radiative equilibrium 
codes to be extended to three dimensions and we are developing such 
techniques (Bjorkman \& Wood 2001).

\acknowledgements
We acknowledge financial support from NASA's Long Term Space Astrophysics 
Research Program, NAG5~6039 (KW), NAG5~8412 (BW); the National Science 
Foundation, AST~9909966 (BW and KW); the HST Archival Research Program 
AR-08367.01-97A, (BW, KW, KS); NASA's Astrophysical Data Program, 
NAG5~3904 (MJW).  KSB is a Cottrell Scholar of the
Research Corporation, and gratefully acknowledges their support.  
DSS was funded by the NSF Research Experience for Undergraduates 
program at the Harvard-Smithsonian Center for Astrophysics
and thanks Christine Jones, Jonathan McDowell, and Tania Ruiz
for the generous donation of time and self to make the
SAO Intern Program as rewarding as it has been.
We thank Karl Stapelfeldt for providing the WFPC2 image of T~Tau and 
Ted von Hippel for assistance with the WIYN queue observations.

%%%%%%%%%%%%%%%%% figures and tables %%%%%%%%%%%%%%%%%%%%

\begin{deluxetable}{lcccccc}
\tablenum{1}
\tablewidth{0pt}
\tablecaption{Parameters for Dust Grains}
\tablehead{
\colhead{Wave Band} & \colhead{$\kappa$ (cm$^2$/g)} &
\colhead{$a$} & \colhead{$g$} & \colhead{$P$}
}
\startdata
%& & & & & & \\
$V\dotfill$  & 219 &  0.54 & 0.44 & 0.43 \\
$I\dotfill$  & 105 &  0.50 & 0.35 & 0.51 \\
$J\dotfill$  & 65 &  0.46 & 0.32 & 0.58 \\
$H\dotfill$  & 38 &  0.42 & 0.29 & 0.59 \\
$K\dotfill$  & 22 &  0.36 & 0.25 & 0.60 \\
\enddata
\end{deluxetable}

%------------------------------FIGURE -------------------
\begin{figure}[t]
\centerline{\plotfiddle{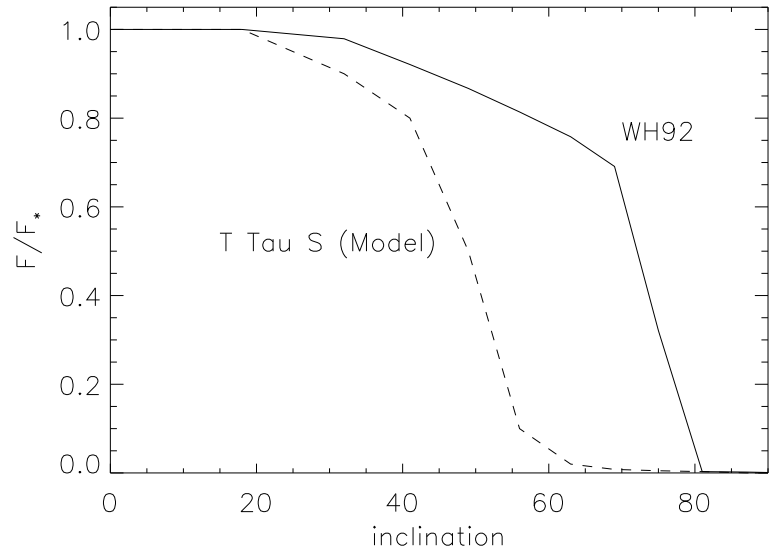}{6in}{0}{65}{65}{-420}{-20}}
\caption{Solid line: angular dependence of emerging flux from a T~Tauri 
star+disk system as determined by Whitney \& Hartmann (1992).  Dashed line: 
the more polar-concentrated angular distribution adopted for our modeling 
of T~Tau~S in Fig.~2d.}
\end{figure}

%------------------------------FIGURE -------------------
\begin{figure}[t]
\centerline{\plotfiddle{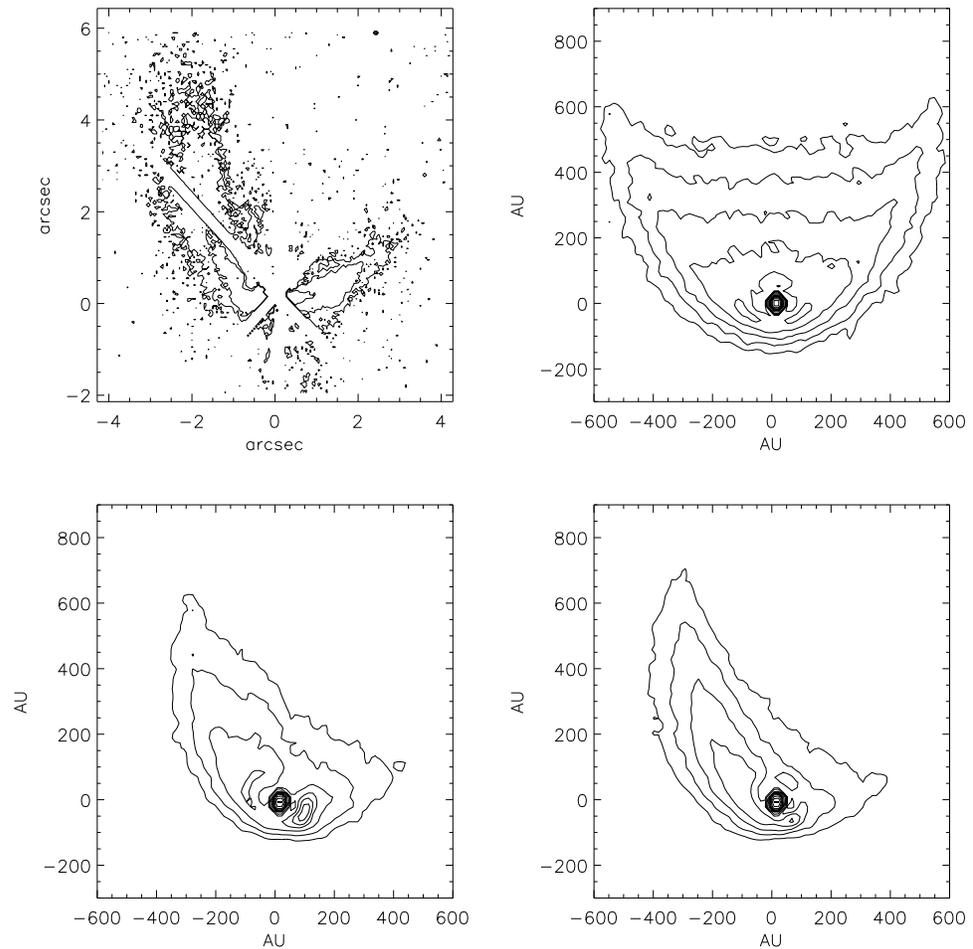}{6in}{0}{65}{65}{-420}{-20}}
\caption{(a) WFPC2 $V$ band (F555W) image of T~Tau.  (b) Axisymmetric, 
single source scattered light model of Stapelfeldt et al. (1998). 
(c) Scattered light model adopting axisymmetric envelope+cavities, but 
with illumination from WH92 star+disks (Fig.~1).  The primary is relatively 
unobscured and its disk rotation axis is aligned with the rotation axis of 
the TSC envelope.  The secondary is oriented so that it is edge-on to our 
line of sight and its rotation axis points towards the upper left corner 
in this figure, breaking the axisymmetry of the illumination and yielding the 
asymmetric scattered light image.  (d) As for (c), but the secondary has 
a more extreme angular distribution (see Fig.~1) yielding a larger 
asymmetry in the scattered light image.  The images in Figs.~2c and 2d have 
been rotated clockwise by $20^\circ$, all models are viewd at $i=40^\circ$.  
In all panels the contour spacing is at half-magnitude intervals.  }
\end{figure}

%------------------------------FIGURE -------------------
\begin{figure}[t]
\centerline{\plotfiddle{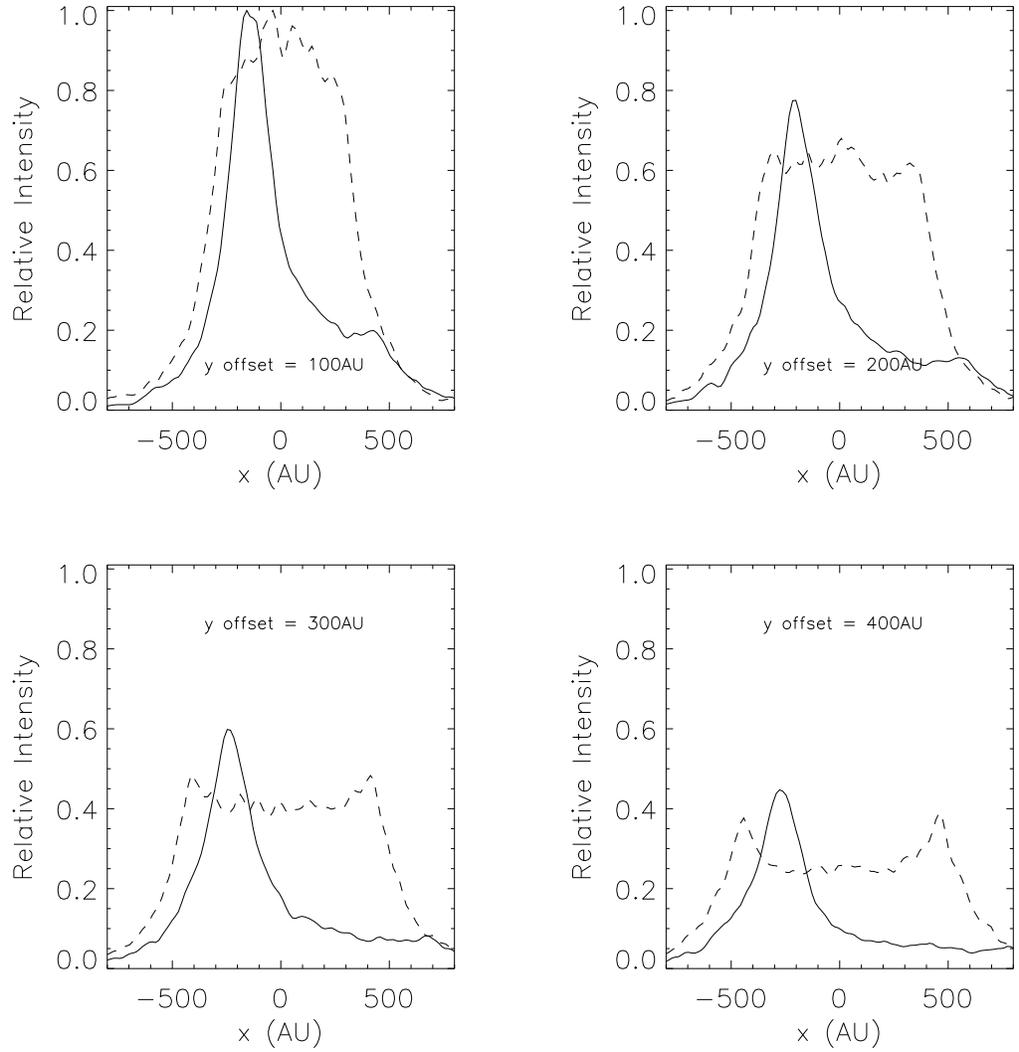}{6in}{0}{65}{65}{-420}{-20}}
\caption{Horizontal cuts at various offsets across the scattered light 
images from Fig.~2b (axisymmetric model, dashed line) and Fig.~2d 
(asymmetric illumination model, solid line).  Note the larger intensity 
contrast of the asymmetric illumination model.}
\end{figure}

%------------------------------FIGURE -------------------
\begin{figure}[t]
\centerline{\plotfiddle{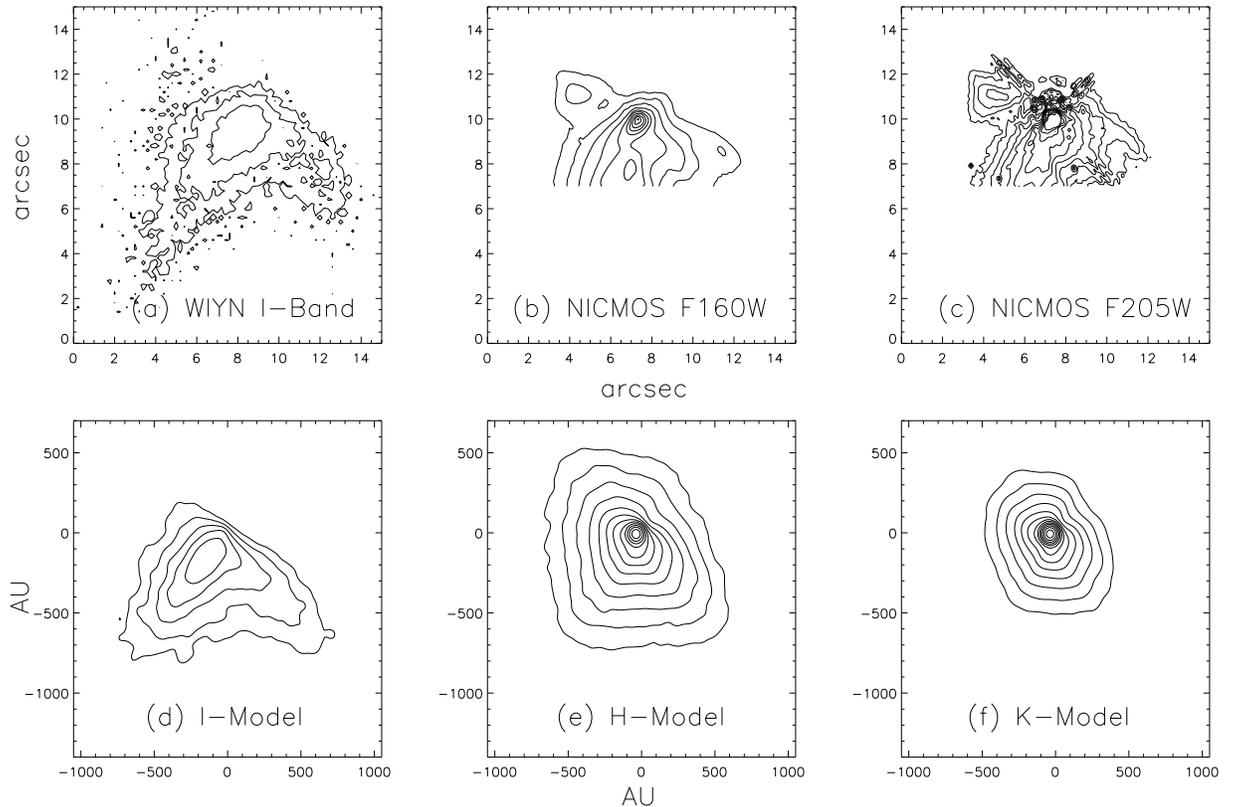}{6in}{90}{65}{65}{20}{100}}
\caption{Scattered light images and models for IRAS~04016+2610.  All 
panels have contours spaced at half-magnitude intervals and the image 
and model panels are the same physical size, adopting a distance to 
IRAS~04016+2610 of 140pc so $1''=140$~AU.  (a) WIYN 
$I$ band image.  (b) NICMOS F160W image.  (c) NICMOS F205W image.  Panels 
(d), (e), and (f) show our scattered light models at $I$, $H$, and $K$ 
bands for a TSC envelope with {\it two} sets of cylindrical bipolar outflow 
cavities (see text and Fig.~5).  In all panels the 
contour spacing is at half-magnitude intervals.  
The scattered light images have been 
rotated clockwise by $220^\circ$ so that the scattered light pattern formed 
by the primary cavity (aligned with the TSC rotation axis) is aligned with 
the primary outflow cavity identified labeled by 
Lucas et al. (2000, Fig.~2).  The models are viewed at an inclination of 
$i=60^\circ$ from the primary outflow cavity.}
\end{figure}

%------------------------------FIGURE -------------------
\begin{figure}[t]
\centerline{\plotfiddle{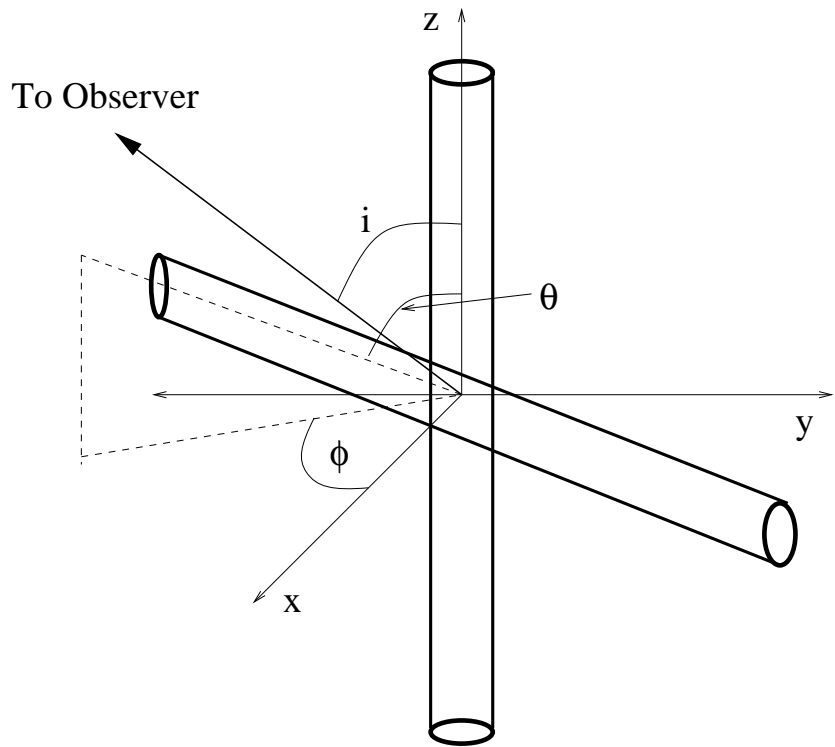}{6in}{0}{65}{65}{-450}{-20}}
\caption{Schematic of the outflow cavity geometry and orientation used in 
our model of IRAS~04016+2610 (Fig.~4).}
\end{figure}

\end{document}